\documentclass[aip,jap,reprint,groupaddress,noshowpacs]{revtex4-1}
\pdfoutput=1

\usepackage[colorlinks=false,bookmarks=true]{hyper ref}
 
\usepackage{mathrsfs}
\usepackage{epsfig}
\usepackage{amsmath}
\usepackage{subfigure}
\usepackage{amsfonts}
\usepackage{lineno}

\usepackage{graphicx}
\usepackage{amssymb}
\usepackage{longtable}

\begin{document}

\title{Focal mechanism estimation by classification}

\author{Ben G. Lasscock}
\affiliation{ION Geophysical, Houston, TX, USA}
\author{Brendon J. Hall}
\affiliation{ION Geophysical, Houston, TX, USA}
\author{Michael E. Glinsky}
\affiliation{Geotrace Technologies, Houston, TX, USA}

\begin{abstract}
A classification technique for identifying focal mechanism type and
fault plane orientation based on the polarity of P-wave ``first
motion'' data is derived. A support vector machine is used to classify
the polarity data in the space of spherical harmonic functions. The
classification is non-parametric in the sense that there is no
requirement to make {\it a priori} assumptions source mechanism. A
metric of similarity potentially able to distinguish shear versus
tensile dislocation without requiring estimation of the fault plane
orientation is a natural consequence of this procedure. Going further,
correlation functions between template source mechanism is derived, gives
an estimate of fault plane orientation assuming a particular source mechanism.
\end{abstract}

\maketitle

\section{Introduction}

In this article we discuss a method to robustly classify the focal
mechanism of a seismic event and estimate fault plane orientation
using observations of the polarity of P-wave ``first motion'' data.
The input of this problem is a set of observations of displacement due
to a seismic event across a seismic array (for microseismic
monitoring) or seismic network.  

Existing methods for focal mechanism classification, such as HASH
\citep{Hardebeck:02} are based on a least squared optimization given a
specific hypothesis of a double-couple type event. However, the least
squares optimization can be unstable where the first motion polarity
is misclassified. To stabilize this, the HASH algorithm implements an
iterative scheme and returns the average over a set of acceptable
solutions as the preferred mechanism. Where identifying acceptable solutions 
requires strategies for outlier rejection.

By comparison, the method we propose identifies the most likely solution
for the nodal lines of these radiation patterns in a basis of
spherical harmonic functions using classification. Since the spherical
harmonics are the eigenfunctions of the scalar wave-equation, these
solutions for the nodal lines can then be used to identify the focal
mechanism of the event. As such, this approach is entirely
non-parametric in the sense that we do not need to presuppose-suppose the
type of focal mechanism ahead of time. 
One advantage of this approach is that it admits a natural metric of
similarity between the focal mechanism types (e.g. double couple,
CLVD) without requiring an estimate of source type or fault plane
orientation.
A procedure for computing the correlation between template focal
mechanisms types and the non-parametric estimate is derived, along
with a corresponding estimate of the fault plane orientation.

Finally, we show that the performance of the classification approach compared to
HASH using the Northridge data-set supplied with the HASH software \citep{HASH:08}.

\section{Mathematical Foundations}

It is well known complex seismic sources can be represented by systems
of force couples \citep{Aki:02,Menahem:81} and that angular variation of
the displacement due to a seismic event can be used to characterize
the source mechanism and its orientation. 
We limit our analysis to the simplified case of a compressional (or
P-wave) arrival in a homogeneous isotropic media. 
Our approach will be to encode the angular dependence of the displacement
in the basis of spherical harmonics. It will be shown that by working in this basis,
event classification (e.g. shear versus tensile failure) and estimation of the 
fault plane orientation is mathematically convenient.

Physically, the spherical harmonic functions are also eigenfunctions
of the scalar wave-equation.  Solutions to the angular dependence of
the P- and S-wave radiation patterns are tabulated
in \citet{Menahem:81}.
Given a particular orientation of the fault plane, the angular
dependence of displacement due to standard focal mechanisms such as
the double couple and clvd sources are also known in this basis. As
such we can relate solutions to the scalar wave equation, to particular 
source mechanism.

But more generally, any square-integrable function, defined on the
2-sphere, can be expressed in terms of the orthonormal basis:
\begin{eqnarray}
\label{eq:ortho}
f(\vec{x}) &=& \sum_{l=0}^{\infty}\sum_{m=-l}^{l}\hat{f}_{lm}Y_{lm}(\theta,\phi)\ ,
\end{eqnarray}
where $Y_{lm}(\theta,\phi)$ are the spherical harmonic functions of
azimuth $\phi$ and inclination $\theta$, and $\hat{f}_{lm}$ are complex
coefficients. The variable indices by $l$- and $m$-respectively are
commonly referred to as the degree and order of the basis function. It
is also common to refer to each degree $l=0,1,2$ as the monopoles,
dipole and quadrupole terms (in deference to the multipole expansion
of electro-statics).
Where this series has good convergence properties, we can view this 
expansion as a kind of data compression, in the sense that a compact
set of coefficients $\hat{f}_{lm}$ provides a concise description 
of the function.
An important mathematical property of this expansion is that the norm of the
coefficients in the span of each degree are rotationally invariant. 
As such a metric for distinguishing the type of seismic event, 
for example shear-failure versus tensile dislocation,
exist without knowledge of the orientation of the respective
fault planes. 
Template solutions for common focal mechanisms, given a particular
orientation, are known in this basis in Table~\ref{tab:theory}. This table
provides a theoretical template in terms of the relative weighting of
basis functions for common seismic events. With more general solutions
tabulated in \citet{Menahem:81}.
Hence the basis of spherical harmonics, at least in the isotropic
media, diagonalizes the problem of characterizing standard seismic
source mechanisms.

\subsection{Kernel methods}
\label{ssec:kernel_methods}

In this section, we will show how modern techniques of classification
taken from the discipline of machine learning can be applied to
earthquake seismology. 

Conditional on a set of event picks of a P-wave arrival and a velocity
model, we can describe the inputs our problem as the set
$\{\vec{x}_{i},y_{i}\}$ of coordinates on a unit sphere (parameterized
by a azimuth and take-off angle) and the sign of the first motion of an
incident 'P-wave' seismic signal. The input data $y_{i}$ is from one
of two classes, positive first motion polarity, or negative first
motion polarity. From this set of inputs our goal is to estimate the
nodal lines of the radiation pattern.

Since the data is imperfect, it is expected that there maybe mis-classification error. 
As such we need a classification algorithm that is robust to a degree of mis-classification 
of the first motion. Further, we require a classification algorithm that is tractable 
on the surface of a sphere and provides a convenient measure of uncertainty. 

The support vector machine \citep{Smola:02} solves for the maximum
separating hyper-plane (nodal lines) between linearly separable
classes of data, with the width of the hyper-plane providing a natural metric
of uncertainty.  Soft thresholding extensions of this support vector
machine \citep{Smola:02} provides robustness to mis-classification.
We elaborate further on the support vector classifier in Appendix~\ref{app:svm}.
Based on a set of input training data, the support vector machine learns the function:
\begin{eqnarray}
f(\vec{x}) = \sum_{i=1}^{N} \alpha_{i}y_{i}k(\vec{x},\vec{x}_{i}) + \beta_{0}\ ,
\end{eqnarray}
by constrained optimization. Here, $y_{i}$ is the polarity of the
$i^{th}$ arrival, the coefficients $\alpha_{i}$ are Lagrange
multipliers that are solved for, an output of the algorithm,
$\beta_{0}$ is a constant and $k(\vec{x},\vec{x}_{i})$ is a kernel.
Domains in the input space are defined by the sign of the function
$f(\vec{x})$, i.e. the solution $f(\vec{x})$ parameterizes the nodal
lines.  The kernel function $k(\vec{x},\vec{x}_{i})$ can be thought of
as a generating functional projecting the classifier into a higher
dimensional feature space.  The so-called class of dot-product kernels are
appropriate for classification on a sphere.
We will show that a particular kernel:
\begin{eqnarray}
\label{eq:kernel}
k(\vec{x},\vec{x}_{i}) &=& (\langle \vec{x}, \vec{x}_{i} \rangle + 1)^{d}\ .
\end{eqnarray}
projects into a basis of spherical harmonic functions, truncating at degree-$d$.

Expanding the kernel in Eq.~\eqref{eq:kernel} in a basis of Legendre polynomials $P_{l}$:
\begin{eqnarray}
f(\vec{x}) &=& \sum_{i=1}^{N}\alpha_{i}y_{i}\sum_{l=1}^{\infty}a_{l}P_{l}(\langle \vec{x}, \vec{x}_{i} \rangle)\ ,
\end{eqnarray}
where the coefficients of the expansion are:
\begin{eqnarray}
a_{l} &=& \int_{-1}^{1}\ dx\ (x + 1)^{d} P_{l}(x)\ .
\end{eqnarray}
This integral evaluates to \cite{Smola:02}:
\begin{align*}
 a_{l} &=
  \begin{cases}
   \frac{2^{d+1}\Gamma(d+1)}{\Gamma(d+2+l)\Gamma(d+1-l)} + \frac{1}{2}\sqrt{\frac{1}{\pi}}\beta_{0}\delta_{l0}  & \text{if } l \leq d \\
   0       & \text{otherwise}
  \end{cases}\ ,
\end{align*}
here we have absorbed the constant term into the coefficient for $l=0$.
Hence the parameter $d$ in the definition of the kernel truncates the 
expansion in Legendre polynomials at degree $d$, which is a useful feature.
To formulate an expression in terms of the spherical harmonic functions we use the addition theorem:
\begin{eqnarray}
P_{l}(\langle\vec{x},\vec{x}_{i}\rangle) &=& \sum_{m=-l}^{l} Y^{*}_{lm}(\theta^{\prime},\psi^{\prime})Y_{lm}(\theta,\psi)\cr
f(\vec{x}) &=& \sum_{i=1}^{N}\alpha_{i}y_{i}\sum_{l=1}^{\infty}a_{l}\
    \frac{4\pi}{2l + 1} \times \cr
    & &\hspace{2mm}\sum_{m=-l}^{l} Y^{*}_{lm}(\theta^{\prime},\psi^{\prime})Y_{lm}(\theta,\psi)\ ,
\end{eqnarray}
where $Y^{*}_{lm}$ notation denotes the complex conjugate of the spherical harmonic function.
Collecting terms:
\begin{eqnarray}
f(\vec{x}) &=& \sum_{l=1}^{\infty}\sum_{m=-l}^{l} \hat{f}_{lm}Y_{lm}(\theta,\psi)\cr
\hat{f}_{lm} &=& \frac{4\pi}{2l + 1}\sum_{i=1}^{N}\alpha_{i}y_{i}a_{l}Y^{*}_{lm}(\theta^{\prime},\psi^{\prime})\ ,
\end{eqnarray}
each of these coefficients maps the separating margin
onto the basis of spherical harmonics, from which we can assign
physical meaning.

\subsection{Assumptions of symmetry}

Physically, seismic events such as earthquakes can be modeled as a
closed system.  As such conservation of momentum, and angular
momentum can be enforced. For the problem of earthquake
classification, assuming an isotropic media, this is equivalent to the
requirement that the solution be even under parity transformation.  To
ensure that this is the case we map observations from the upper to
lower half spheres (and vice verse) to ensure that the optimal
non-parametric solutions respect this symmetry.

%For the problem of microseismic monitoring, seismic events are currenlty 
%thought to be due shear failure of the material due to small perturbations 
%of the stress field created by hydro-fraccing\cite{}. In this picture, 
%the external applied forces due to hydro-fraccing are  seen to be small compared 
%to the internal stresses within the system, and so the solutions should 
%still transform evenly under partity.

\subsection{Mapping the kernel estimation to a parsimonious solution}

The kernel estimation of the nodal lines is entirely driven by the data.
However, analysts require a mapping onto a set of parsimonious solutions from which 
they can interpret physical meaning. In this section we derive a rotationally invariant
signature of a seismic event based on its multipole expansion. We will then go on 
to develop formalism for estimating correlation functions between events with a
nested optimization over the fault plane orientation.

Seismic sources characterized as systems of force couples are orientated
with respect to the normal of a planar fault and the direction of the
slip. Particular solutions for the angular dependence can be derived
given a particular orientation \citep{Menahem:81}; template solutions
for standard seismic sources are tabulated in
Table \ref{tab:theory}. A signature of the seismic event can be
expressed in terms of the norm of the coefficients at each order:
\begin{eqnarray}
\label{eq:signature}
q_{l} = \sum_{m=-l}^{l} \hat{f}_{lm}^{*}\hat{f}_{lm}\ ,
\end{eqnarray}
where each $q_{l}$ tells us the relative contribution of monopole,
dipole, quadrupole, etc. composition of the source. In Appendix~\ref{app:opt} 
we show that this signature is invariant under rotations.

%
%To estimate the orientation of the focal plane, define
%the correlation function between an observed and template solution $f$
%and $g$ as the inner product:
%
Going further, we can define a correlation function between seismic events as:
\begin{eqnarray}
\langle g, f\rangle = \int d^{3}x g^{*}(R(\alpha,\beta,\gamma) \cdot \vec{x}) f(\vec{x}) 
\end{eqnarray}
where $R(\alpha,\beta,\gamma)$ is a rotation matrix parameterized by
the Euler angles. The role of the rotation matrix in the correlation
function is to recognize that the relative orientation of the focal
plane between the two events maybe different. The action of the
rotation applied in the input space generates a rotation in the
feature space.  The set of matrices that perform this rotation in the
feature space of spherical harmonics are called Wigner's D matrices,
see \citet{Morrison:87} for review. In Appendix~\ref{app:opt} we show
that we can write the rotation of the estimated nodal lines as:
\begin{eqnarray*}
f(R(\alpha,\beta,\gamma) \cdot \vec{x}) &=& \sum_{l=0}^{\infty}\sum_{m,n=-l}^{l} D^{l}_{mn}(\alpha,\beta,\gamma)\hat{f}_{ln}Y_{lm}(\theta,\phi)\ .
\end{eqnarray*}
Using the orthogonality condition of the spherical harmonics leads to a discrete expression for the correlation function:
\begin{eqnarray}
\label{eq:corr}
\langle g, f\rangle &=& \sum_{l=0}^{\infty}\sum_{m,n=-l}^{l}D^{l*}_{mn}(\alpha,\beta,\gamma)\hat{g}^{*}_{ln}\hat{f}_{lm}\ ,
\end{eqnarray}
where we have made use of the fact that the D-matrices do not mix coefficients of different degree.
By optimizing the correlation function over the Euler angles we obtain the 
correlation between and two source mechanisms, and corresponding orientation 
of the fault plane. Details of the optimization algorithm can be found in Appendix~\ref{app:opt}.

\section{Application to Northridge data}

In this section we apply the support vector classifier to the
Northridge dataset supplied with the USGS HASH
software\footnote{http://earthquake.usgs.gov/research/software/index.php}.
The HASH algorithm \citep{HASH:08} provides focal mechanism
classification using first motion polarity based on least squares,
and is considered to be robust.

The example dataset north1 from \citet{HASH:08} is analyzed by HASH and
the support vector classifier. The default parameterization of HASH is
used, except that we extend the number of iterations to from 30 to
3000.  The north1 dataset contains the azimuth, take-off angles and
first arrival polarity for a collection of earthquakes from the
Northridge region of California, recorded by the Southern California
Seismic Network. Polarity reversals are applied where appropriate, see
the HASH manual \citep{HASH:08}. 

\begin{figure}
\begin{tabular}{c}
\includegraphics[height=6.5cm,width=6.5cm]{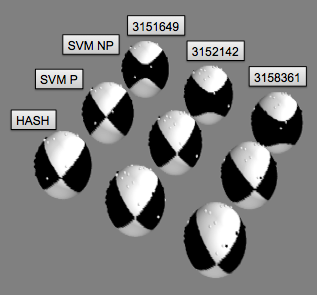}
\end{tabular}
\caption{\label{fig:demo} Shows the HASH (front row), (P) parametric
  SVM equivalent solution (middle row) and (NP) non-parametric
  solution (back row). The catalog event ids are 3151649, 3152142
  and 3158361.  }
\end{figure}

In Fig.~\ref{fig:demo} we show a graphical example of focal mechanism solutions derived from 
HASH compared to the parametric and non-parametric estimates derived from the classifier. Like 
HASH, the parametric-solution assumes {\it a priori} a double-couple (shear) event. Given this 
template for the event in the basis of spherical harmonic functions Table~\ref{tab:theory}, the fault 
plane orientation is estimated using Eq.~\eqref{eq:corr}. Whereas the non-parametric solution is simply 
the superposition derived from Eq.~\eqref{eq:ortho} (without assuming a particular type of mechanism or 
solving for the fault plane orientation).

Extending the analysis to the entire Northridge (north1) dataset, in
Table \ref{tab:results}, we evaluate the respective algorithms by their overall 
rate of mis-classification. 
\begin{table}
\caption{\label{tab:results}
The average rate of misclassification for the Northridge (north1) dataset.
The parametric and non-parametric classifiers are compared to the HASH 
algorithm.
}
\begin{tabular}{ccc}
svm parametric & svm non-parametric & HASH \cr
\hline
15.5\% & 14.7\% & 15.3\%
\end{tabular}
\end{table}

\section{Conclusions}

We presented a new technique for classifying the focal mechanism
classification using P-wave first motion polarity data.  This new
approach provides classification for a general set of sources
including combinations of shear- and tensile failure (double couple
and CLVD). Robustness to misclassification error is achieved using
soft thresholding, rather employing importance weighting observations
or by heuristic outlier rejection. This suits the problem at hand
where the observations along the nodal lines are the most informative,
but also the most likely to be misclassified.
We also derive a rotationally invariant source signature that may
provide comparison between seismic events without requiring a
parameterization of the source type or fault plane orientation.  This
signature has applications for distinguishing shear- from tensile
failure, or comparing linear combinations of both.

Finally we compare the performance of the classifier to the Northridge 
data-set supplied with HASH \citep{HASH:08} and performs as well as the 
current state of the art.

\appendix

\section{The support vector classifier}
\label{app:svm}

In this section we provide a short overview of the support vector classifier. 
We follow \citet{Tibshirani:01} and \citet{Smola:02} as references. The software 
implementation of the support vector classifier used in this report 
sklearn version 0.14\footnote{http://scikit-learn.org}.

First motion polarity on a 2-sphere is not linearly
separable in a rectangular coordinate system, therefore the strategy
of the support vector machine is to linearize the problem by expanding into 
a higher dimensional space. The space where the observations are made is
commonly referred to as the input space, the space where the
classification is performed called the feature space. The efficiency
of the support vector machine is subtended by the so-called kernel
trick, which allows this expansion to into feature space to be carried
out in terms of inner products calculated in the input space.

Using the notation from Sec.~\ref{ssec:kernel_methods} each datum is
described a Cartesian coordinate and a class
($\{\vec{x}_{i},y_{i}\}$), where the the class $y_{i}$ is the first
motion polarity.
Consider the example where we have $N$ observations of the first
motion polarity over the surface of a unit sphere. Suppose that there
is some mapping $\vec{\phi}(\vec{x})$ into a higher dimensional space
where this can be considered as a linear problem. That is, where there
exists some hyper-plane separates the two classes:
\begin{eqnarray}
f(\vec{x}) = \langle\vec{\beta},\vec{\phi}(\vec{x})\rangle + \beta_{0}\ .
\end{eqnarray}
The strategy the support vector classifier uses is to optimize the
width of a margin separating classes in this domain, which can be
written as the optimization:
\begin{eqnarray}
\min_{\vec{\beta},\vec{\eta}} \frac{1}{2}\Vert{\beta}^{2}\Vert + C \sum_{i=1}^{N}\eta_{i}\ ,
\end{eqnarray} 
subject to the constraints:
\begin{eqnarray}
y_{i}(\langle\vec{\beta},\vec{\phi}(\vec{x}_{i})\rangle + \beta_{0}) &\geq& 1 - \eta_{i}\cr
\eta_{i} &\geq& 0,\ i=1,...,N\ .
\end{eqnarray} 
The addition of a set of slack variables $\eta_{i}$ stabilizes the
algorithm by allowing for for mis-classification, but with a
penalty. The penalty leads the optimization to prefer
mis-classification close to the nodal line, which it is expected to be
most prevalent.

This makes the algorithm robust where the data is not exactly separable in the space spanned by $\vec{\phi}$, 
a problem which will be caused by mis-classification of the first motion polarity. 

The optimization described above is problem of quadratic programming, which is 
solved by introducing a set of Lagrange multipliers $\alpha_{i}$ for each constraint.
The so-called dual form of the Lagrangian is:
\begin{eqnarray}
\mathcal{L} = \sum_{i=1}^{N}\alpha_{i} - \frac{1}{2}\sum_{i,j=1}^{N}\alpha_{i}y_{i}\alpha_{j}y_{j}\langle\vec{\phi}(\vec{x_{i}}), \vec{\phi}(\vec{x_{j}})\rangle\ .
\end{eqnarray} 
For a certain class of function $\vec{\phi}(\vec{x})$, the inner product (called a kernel) maybe evaluated in the input space directly. 
As an example, we show that the inner product kernel Eq.~\eqref{eq:kernel} maps to an expansion in spherical harmonic functions, a 
natural basis for our problem.  For a separable dataset, the Lagrange multipliers are non-zero only for points along the optimal separating 
hyper-plane. The coordinates of the corresponding data are called support vectors.

\section{Optimization for estimating fault plane orientation}
\label{app:opt}

In Sec.~\ref{ssec:kernel_methods} we have established that a set of nodal lines along a unit sphere 
can be expressed as a series of spherical harmonic functions:
\begin{eqnarray}
f(\vec{x}) = \sum_{l=0}^{\infty}\sum_{m=-l}^{l}\hat{f}_{lm}Y_{lm}(\theta,\phi)\ .
\end{eqnarray}
Were we to perform a rotation of this coordinate system, the same
general expression must still hold, and the norm of the coefficients
must be invariant, however the individual coefficients themselves may
change.  An irreducible representation of the rotations group in the
span of spherical harmonic functions is given by the Wigner-D
matrix. That is, for some rotation $R(\alpha,\beta,\gamma)$:
\begin{eqnarray}
f(R\cdot\vec{x}) = \sum_{l=0}^{\infty}\sum_{m=-l}^{l} D^{l}_{mm'}\hat{f}_{lm'}Y_{lm}(\theta,\phi)\ .
\end{eqnarray}
The properties of this representation of rotations is that it is
unitary and irreducible in the span of the harmonic functions at
each order.  These properties ensure that the norm of the
coefficients of the expansion, at each degree in the basis of harmonic
functions is invariant under rotations:
\begin{eqnarray}
q_{l} &=& \sum_{m=-l}^{l} \hat{f}^{*}_{lm}\hat{f}_{lm} \cr
&\rightarrow&  \sum_{m,m'=-l}^{l} \hat{f}^{*}_{lm} D^{l *}_{mm'}D^{l}_{m'm}\hat{f}_{lm}\cr
&=& \sum_{mm'=-l}^{l} \hat{f}^{*}_{lm} \delta_{m'm}\hat{f}_{lm}\cr
&=& q_{l}\ .
\end{eqnarray}
An explicit form of Wigner-D matrices is:
\begin{eqnarray}
D^{l}_{mn}(\alpha,\beta,\gamma) &=& {\rm e}^{-im\alpha}d^{l}_{mn}(\beta){\rm e}^{-in\gamma}\ ,
\end{eqnarray}
where the so-call ``little Wigner-d'' matrix is purely real. We use the formulation 
of \citet{Morrison:87} for the Wigner-D matrices, except for an overall sign difference 
in the $d^{2}_{10}$ term.

Next we apply these rotation matrices to optimize the
correlation function in Eq.~\eqref{eq:corr}.  The discrete form of the
correlation function is derived by using the orthogonality condition
(suppressing the argument $\alpha$, $\beta$ and $\gamma$):
\begin{eqnarray*}
\langle g, f\rangle &=& \int d^{3}x\ g^{*}(R\vec{x}) \  f(\vec{x}) \cr
&=& \sum_{l,l'=0}^{\infty}\sum_{n,m'}\int d\Omega\ D^{l*}_{m'n}\hat{g}^{*}_{l'n}\hat{f}_{lm}Y_{l'm'}^{*}Y_{lm}\cr
&=& \sum_{l,l'=0}^{\infty}\sum_{m,m',n} D^{l*}_{m'n}\hat{g}^{*}_{l'n}\hat{f}_{lm}\delta_{ll'}\delta_{mm'}\cr
&=& \sum_{l=0}^{\infty}\sum_{m,n} D^{l*}_{mn}\hat{g}^{*}_{ln}\hat{f}_{lm}\ .
\end{eqnarray*}
The optimization with respect to the variables $\alpha$, $\beta$ and
$\gamma$ is performed numerically.
%using a truncated Newton method,
%with software implementation of this method taken from the scipy
%minimize library\footnote{http://docs.scipy.org}.
%
%To evaluate the Jacobian and the Hessian 
%of this correlation function, it is convenient to use the change of variables \cite{Kovacs:02}:
%$R(\xi,\eta,\omega) = R_{1}(\xi,\frac{\pi}{2},0)R_{2}(\eta,\frac{\pi}{2},\omega)$ with:
%\begin{eqnarray*}
%\xi = \alpha - \frac{\pi}{2},\ \eta = \pi - \beta,\ \omega = \gamma - \frac{\pi}{2}\ .
%\end{eqnarray*}
%This change of variables transforms the discrete form of the correlation function to:
%\begin{eqnarray*}
%D^{l}_{mn}(\xi,\eta,\omega) &=& D^{l}_{mn}(\xi,\frac{\pi}{2},0)D^{l}_{mn}(\eta,\frac{\pi}{2},\omega)\cr
%\langle g, f\rangle &=& \sum_{l=0}^{\infty}\sum_{m,n,m'} d^{l}_{mn}(\frac{\pi}{2})d^{l}_{nm'}(\frac{\pi}{2})\hat{g}^{*}_{%l'm'}\hat{f}_{lm}\times \cr
%& &\hspace{3mm} {\rm e}^{-i(m\xi + n\eta + m'\omega)}\ .
%\end{eqnarray*}
%In this basis the so-called ``little'' Wigner-d matrix is real and
%constant, and hence the general expression for the derivatives of the
%correlation function are trivial to compute analytically.

\section{Seismic source templates}

In this section we provide a lookup table of theoretical templates of
the compression mode of the displacement oriented radially, for
standard seismic sources in a homogeneous isotropic media. 
Solutions in terms of the Hansen vectors are taken from Table 4.4 of
\citet{Menahem:81}. The solutions for the first arrival are given 
in terms of the Hansen vector $\vec{L}$ (in spherical polar coordinates) of the form:
\begin{eqnarray}
\vec{L}_{lm}(r,\theta,\phi) &=& \vec{\nabla} h^{2}_{l}(r) \tilde{Y}_{lm}(\theta,\phi) \ ,
\end{eqnarray}
where $h^{2}_{l}(r)$ is the spherical Hankel functions of a second kind. The amplitudes of the 
first break are required to be measured radially, the projection of the Hansen vector radially is:
\begin{eqnarray}
\hat{r}\cdot \vec{L}_{lm}(r,\theta,\phi) &=& \frac{\partial}{\partial r} h^{2}_{l}(r) \tilde{Y}_{lm}(\theta,\phi) \ ,
\end{eqnarray}
where $\hat{r}$ is the radial unit vector.
Asymptotically, the Hankel functions tend to \citet{Morse:53}: 
\begin{eqnarray}
h_{l}^{2}(x) &=& \frac{1}{x}(i)^{l+1}{\rm e}^{-ix}\ ,
\end{eqnarray}
which introduces a relative sign when collecting terms of degree $0$ and $2$.

We also note that the normalization of the spherical harmonics used in \citet{Menahem:81}, to our definition:
\begin{eqnarray}
\tilde{Y}_{lm}(\theta,\phi) &=& (-1)^{m}\sqrt{\frac{4\pi(l+m)!}{(2l+1)(l-m)!}} Y_{lm}(\theta,\phi)\ .
\end{eqnarray}

With these adjustments, the amplitudes (up to an overall constant) for
a common set of source mechanism, in terms of the spherical harmonics,
are given in Table~\ref{tab:theory}. The $(\cdot\cdot)$ notation in
this table labels the orientation of the fault normal and direction of
slip respectively.  
\begin{table}
\caption{\label{tab:theory} Describes the angular variation of the
  radial component of displacement in terms of spherical harmonic
  functions. The source templates summarized are double couple (D.C.), 
  tensile dislocation (Tensile) and tangential dislocation (Tangential). 
  The brackets $(\cdot,\cdot)$ define the template
  direction of the fault normal and direction of slip in rectangular
  coordinates.  For the tensile dislocation (CLVD sources) the
  constant $\alpha = 2 + 3\frac{\lambda}{\mu}$, where $\lambda$ and
  $\mu$ are the first Lame parameter and the shear modulus
  respectively.}
\begin{tabular} {ccc}
Source & (Fault normal/slip) & Template \cr
\hline
D.C. &
(31) + (13) & $-i(Y_{12} + Y_{-12})$\cr
Tensile & 
(3) & $\alpha Y_{00} + 4\sqrt{5} Y_{02}$\cr
Tangential & 
(3) & $Y_{02} - \frac{i}{2}(Y_{22} + Y_{-22})$\ .
\end{tabular}
\end{table}


%\bibliography{bib}

\end{document}